\pdfoutput=1

\documentclass[11pt,twoside,a4paper,cmspaper,final,collab]{cms-tdr}
\def\svnVersion{429430P}\def\cmsCernNoTag{CERN-EP-2017-068}\def\cmsCernDate{\today}\def\cmsMessage{Published in Physical Review Letters as \href{http://dx.doi.org/10.1103/PhysRevLett.119.141802}{\doi{10.1103/PhysRevLett.119.141802.}}}

\begin{document}\cmsNoteHeader{HIG-16-027}

\hyphenation{had-ron-i-za-tion}
\hyphenation{cal-or-i-me-ter}
\hyphenation{de-vices}
\RCS$Revision: 429333 $
\RCS$HeadURL: svn+ssh://svn.cern.ch/reps/tdr2/papers/HIG-16-027/trunk/HIG-16-027.tex $
\RCS$Id: HIG-16-027.tex 429333 2017-10-12 22:42:05Z jsalfeld $

\newlength\cmsFigWidth
\ifthenelse{\boolean{cms@external}}{\setlength\cmsFigWidth{0.98\columnwidth}}{\setlength\cmsFigWidth{0.49\textwidth}}
\ifthenelse{\boolean{cms@external}}{\providecommand{\cmsLeft}{upper\xspace}}{\providecommand{\cmsLeft}{left\xspace}}
\ifthenelse{\boolean{cms@external}}{\providecommand{\cmsRight}{lower\xspace}}{\providecommand{\cmsRight}{right\xspace}}
\ifthenelse{\boolean{cms@external}}{\providecommand{\NA}{\ensuremath{\cdots}}}{\providecommand{\NA}{\ensuremath{\text{---}}}}
\ifthenelse{\boolean{cms@external}}{\providecommand{\cmsTableResize[1]}{\relax{#1}}}{\providecommand{\cmsTableResize[1]}{\resizebox{\columnwidth}{!}{#1}}}

\newcommand{\V}{\ensuremath{\mathrm{V}}\xspace}
\newcommand{\W}{\PW}
\newcommand{\ZZ}{\ensuremath{\Z\Z}\xspace}
\newcommand{\mt}{\ensuremath{m_\mathrm{T}}\xspace}
\newcommand{\tZq}{\ensuremath{\PQt\Z\PQq}\xspace}

\cmsNoteHeader{HIG-16-027}

\title{Search for charged Higgs bosons produced via vector boson fusion and decaying into a pair of \texorpdfstring{$\PW$ and $\PZ$}{W and  Z} bosons using proton-proton collisions at \texorpdfstring{$\sqrt{s} = 13$\TeV}{13 TeV}}

\date{\today}

\abstract{A search for charged Higgs bosons produced via vector boson fusion and decaying into $\PW$ and $\PZ$ bosons using proton-proton collisions at $\sqrt{s}=13$\TeV is presented. The data sample corresponds to an integrated luminosity of 15.2\fbinv collected with the CMS detector in 2015 and 2016. The event selection requires three leptons (electrons or muons), two jets with large pseudorapidity separation and high dijet mass, and missing transverse momentum. The observation agrees with the standard model prediction. Limits on the vector boson fusion production cross section times branching fraction for new charged physical states are reported as a function of mass from 200 to 2000\GeV and interpreted in the context of Higgs triplet models.}

\hypersetup{%
pdfauthor={CMS Collaboration},%
pdftitle={Search for charged Higgs bosons produced in vector boson fusion processes and decaying into a pair of W and Z bosons using proton-proton collisions at sqrt(s) = 13 TeV},%
pdfsubject={CMS},%
pdfkeywords={CMS, physics, charged Higgs, WZ fusion, WZ decays}}

\maketitle

The discovery~\cite{ATLAS,CMS} of a Higgs boson~\cite{Higgs1,Higgs2,Higgs3,Higgs4,Higgs5,Higgs6} at the CERN LHC marks an important milestone in the exploration of the electroweak (EW) sector of the standard model (SM)~\cite{SM1,SM2,SM3}. Many aspects of EW interactions at the energy scale of 1 TeV, however, remain to be explored. At the LHC, the study of vector boson scattering (VBS) may reveal hints to extensions of the SM. In particular, extended Higgs sectors with additional SU(2) doublets~\cite{PhysRevD.8.1226,FAYET197414, Craig2012, Branco20121} or triplets~\cite{KONETSCHNY1977433, MAGG198061, PhysRevD.22.2860,PhysRevD.22.2227,CE1,CE2,Chiang:2012cn,Chiang:2013rua} introduce couplings of vector bosons to heavy neutral or charged Higgs bosons.

Searches for charged Higgs bosons ($\PHpm$) at the LHC currently focus on the production and the decay via couplings to fermions~\cite{charged1,charged2,charged3,charged4,charged5,charged6,charged7,charged8,charged9}, well motivated by the minimal supersymmetric standard model~\cite{Dimopoulos:129282}. In this model, the $\PHpm \PQt\PQb$ coupling is the dominant one irrespective of the mass of the charged Higgs boson ($m(\PHpm)$) and $\tan{\beta}$, the ratio of the vacuum expectation values of the two Higgs doublets. Couplings to vector bosons are, however, largely suppressed in these models.

\begin{figure}[htbp]
\centering
\includegraphics[width=\cmsFigWidth]{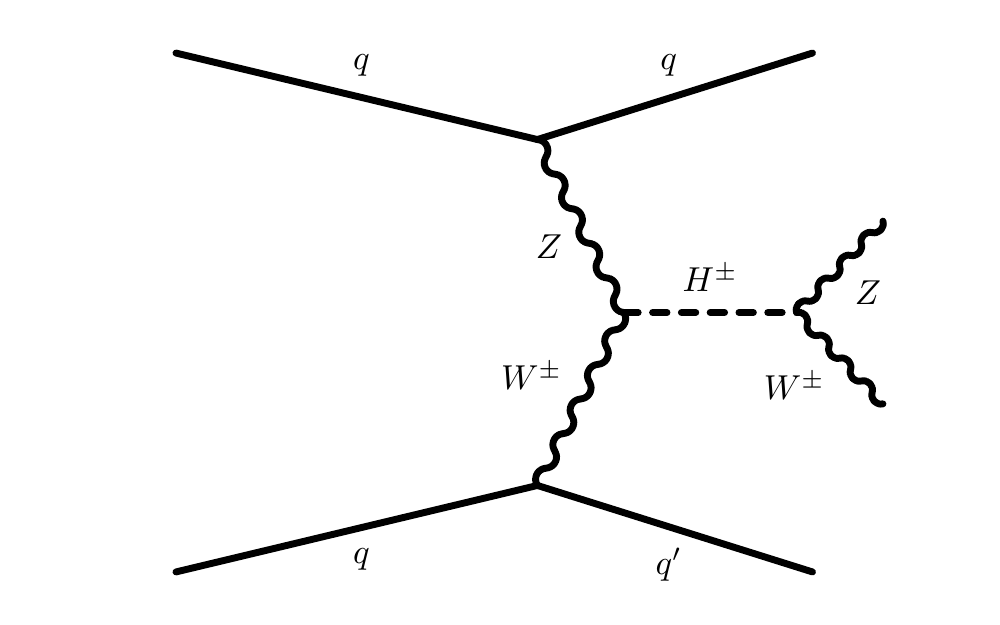}
\caption{Example of a Feynman diagram showing the production of charged Higgs bosons via VBF.}
\label{fig:Feyn}
\end{figure}

Higgs sectors extended by SU(2) triplets, however, give rise to charged Higgs bosons with couplings to $\W$ and $\Z$ bosons at the tree level. Higgs triplets appear in left-right symmetric~\cite{leftright1,leftright2,leftright3}, little Higgs~\cite{LittleHiggs,Chang:2003zn,Chang:2003un}, and supersymmetric models~\cite{susyGM1,susyGM2} and can generate neutrino masses via the seesaw mechanism~\cite{PhysRevD.22.2860,PhysRevD.22.2227,MAGG198061,seesaw3,seesaw4}. A particularly prominent model is the Georgi--Machacek (GM) model~\cite{GEORGI1985463}, where two SU(2) triplets (one real and one complex) are added to the SM Higgs sector and preserve custodial symmetry for large vacuum expectation values of the SU(2) triplets.
In such models, the charged Higgs bosons are produced via vector boson fusion (VBF) and the couplings depend on $m(\PHpm)$ and the parameter $\sin \theta_{\PH}$, or $s_{\PH}$, where $s_{\PH}^2$ denotes the fraction of the $\W$ boson mass squared generated by the vacuum expectation value of the triplets. A representative Feynman diagram for the production by and decay into a $\W$ and $\Z$ boson pair is shown in Fig.~\ref{fig:Feyn}.

In this Letter, we discuss the search for charged Higgs bosons that are produced via VBF and decay via couplings to $\W$ and $\Z$ bosons. The analysis is performed on a sample of proton-proton collisions collected at $\sqrt{s} = 13\TeV$ center-of-mass energy by the CMS experiment at the LHC. The data sample corresponds to integrated luminosities of 2.3 and 12.9\fbinv recorded during the years 2015 and 2016, respectively. The search is performed using $\W$ and $\Z$ bosons decaying into electrons and muons. The event selection requires two jets with large pseudorapidity separation and a high dijet mass to select a VBF topology. The data are compared to the predictions of the GM model for a charged Higgs boson mass range of $200< m(\PHpm) < 1000$\GeV . In addition, an exclusion limit on the VBF production cross section times branching fraction ($\mathcal{B}$) for $200< m(\PHpm) < 2000$\GeV  is derived. A similar search was performed by the ATLAS Collaboration in proton-proton collisions at $\sqrt{s}= 8$\TeV in the semi-leptonic ($\W\Z \to \PQq\PQq'\ell\ell$) final state~\cite{PhysRevLett.114.231801}. Other experimental constraints on the GM model can be obtained from studies of \PQb-meson decays~\cite{PhysRevD.91.015013} and $\W^{\pm}\W^{\pm}$ VBS processes~\cite{ATLASVBS,CMSVBS}.

The signal samples are produced with \MGvATNLO v2.2.2~\cite{mg_amcnlo}.
$\PW\Z$ production in association with two jets involving exclusively
electroweak interactions at the tree level is generated at leading-order (LO) using \MGvATNLO and is referred to as an EW $\PW\Z$ background. Two-jet-associated $\PW\Z$ production with both the strong and electroweak interaction vertices at the tree level is simulated at next-to-leading order (NLO) using {\POWHEG 2.0}~\cite{Alioli:2008gx,Nason:2004rx,Frixione:2007vw,powheg:2010} and is denoted as a QCD $\PW\Z$ background.
The $\cPZ$+jets, $\Z\gamma$, $\tZq$, $\ttbar \V$, and $\V\V\V$ backgrounds, where $\V$ refers to a $\W$ or $\Z$ boson, are produced at NLO using \MGvATNLO. Simulated $\tZq$ and $\ttbar \V$ events are included in the background referred to as $\V\V\V$. The $\Pg\Pg \to \ZZ$ sample is generated at LO with {\textsc{mcfm}}~\cite{MCFM} and normalized to NLO with a K-factor of 1.7~\cite{Caola:2015psa}. The $\Z\Z$ production via $\Pq\Paq$ annihilation is simulated at NLO with \POWHEG and normalized to the next-to-next-to-leading order (NNLO) cross-section prediction with a K-factor of 1.1~\cite{Cascioli:2014yka}. The \PYTHIA8~\cite{Sjostrand:2015} package is used for parton showering, hadronization, and the underlying event simulation with parameters affecting the underlying event simulation set to the CUETP8M1 tune~\cite{Skands:2014pea,Khachatryan:2015pea}.
The NNPDF~3.0~\cite{nnpdf} set is used as the default set of parton distribution functions (PDFs).
For all processes, the detector response is simulated using a detailed
description of the CMS detector, based on the \GEANTfour~package~\cite{Agostinelli:2002hh}, and event reconstruction is performed with
the same algorithms as used for the data. The simulated samples include additional interactions per bunch crossing (pileup) matching the observed multiplicity in the data of about 11 and 20 interactions per bunch crossing in 2015 and 2016, respectively.

Details of the CMS detector, its performance, and the definition of the coordinate system can be found in Ref.~\cite{CMSdetector}.
The detector features a superconducting solenoid with a diameter of 6\unit{m}, providing a magnetic field of 3.8\unit{T}, and surrounding a silicon pixel and strip tracking detector, a lead tungstate electromagnetic calorimeter, and a brass scintillator hadronic calorimeter. Gas ionization detectors embedded into the steel-flux return yokes, the muon system, are installed around the solenoid. The subdetectors are composed into a barrel and two end cap sections. The hadron forward calorimeter provides calorimetry to pseudorapidities from $\abs{\eta}>3$ to $\abs{\eta}<5$.
A particle-flow technique~\cite{PFT-09-001,PFT-10-001} is employed to identify and reconstruct the individual particles emerging from each collision.

Electrons are reconstructed within $\abs{\eta} < 2.5$. The reconstruction combines the information from clusters of energy deposits in the electromagnetic calorimeter and the trajectory in the tracker~\cite{Khachatryan:2015hwa}.
The selection criteria depend on transverse momentum $\pt$ and $\abs{\eta}$, and on a categorization based on observables sensitive to the amount of bremsstrahlung emitted.
Muons are reconstructed within $\abs{\eta} < 2.4$~\cite{Chatrchyan:2012xi}. The reconstruction combines the information from both the tracker and the muon spectrometer.
Leptons are required to be isolated from other charged and neutral particles in the event. The lepton relative isolation is defined as the ratio of the $\pt$ sum of charged hadrons and neutral particles within a cone of radius $\Delta R = \sqrt{\smash[b]{(\Delta \eta)^{2} + (\Delta \phi)^{2}}} < 0.4$ (where $\phi$ is the azimuthal angle in radians) around the lepton and the lepton $\pt$. The relative isolation, corrected for pileup contributions, is required to be less than 6.5\% (15\%) for electrons (muons). Overall efficiencies of the reconstruction, identification, and isolation requirements for the prompt leptons are measured in the data in several bins of $\pt $ and $ \abs{\eta} $ using a ``tag-and-probe'' technique~\cite{wzxs} applied to a sample of leptonically decaying $\cPZ$ boson events.

Jets are reconstructed using the anti-\kt clustering algorithm~\cite{antikt} with a distance parameter $R=0.4$, as implemented in the \FASTJET package~\cite{Cacciari:fastjet1,Cacciari:fastjet2}, and jet energy corrections are applied~\cite{cmsJEC,cmsJEC2}.
To suppress the top-quark background contribution in its decay to $\PQb$ quarks, the combined secondary vertex \PQb-tagging algorithm~\cite{Chatrchyan:2012jua,CMS-PAS-BTV-15-001} requirement is used, corresponding to an efficiency of about 45\% with a light flavor quark misidentification probability of 0.1\%.

The missing transverse momentum vector \ptvecmiss is defined as the negative vectorial sum of the momenta of all reconstructed particles in an event projected onto the plane perpendicular to the beams, corrected for the pileup contribution~\cite{Cacciari:2007fd}. Its magnitude is referred to as $\ptmiss$.

Events are selected by the trigger system requiring the presence of one or two high $\pt$ electrons or muons.
The trigger efficiency is greater than 99\% for events that pass all
other selection criteria explained in the following.
The selection of events aims to single out three-lepton events with
the VBF topology.
The event selection requires three lepton (electron or muon) candidates that meet the isolation and
identification requirements.
Two leptons are required to have
$\pt > 20\GeV$  and the third lepton is required to have $\pt > 10\GeV$. Events with an additional fourth lepton with $\pt > 10\GeV$ are rejected.
Events are required to have at least two jets with $\pt>30\GeV$, and $\abs{\eta}<4.7$.
The VBF topology is exploited by requiring that the
two jets of highest $\pt$ have a large dijet mass, $m_{jj} > 500\GeV$, and a large pseudorapidity separation, $\left|\Delta \eta_{jj}\right| > 2.5$.
To reconstruct a $\Z$ boson candidate, a pair of same-flavor and opposite-charge leptons is required to have a dilepton invariant mass within 15\GeV of
the nominal $\Z$ boson mass~\cite{PDG2016}.
When there are two or more candidate pairs, the one with
the mass closest to the nominal $\Z$ boson mass is chosen. The
remaining lepton is associated with the $\W$ boson decay, and it is required to have $\pt > 20\GeV$.
The $\ptmiss$ in the event is required to be larger
than 30\GeV to select $\W$ boson decays.
To reject the top-quark background, the event must not have jets passing the $\PQb$-tagging selection.
After these requirements, the signal efficiency is about 10--15\%, depending on $m(\PHpm)$.
For extraction of the signal, the shape of the distribution of the transverse mass variable ($\mt$) obtained from the $\W\Z$ system is used
\begin{equation*}
\begin{aligned}
\mt({\PW\PZ}) = \sqrt{(\ET(\PW) + \ET(\Z))^2 - (\ptvec(\PW)+\ptvec(\Z))^2},
\label{eq:transMass}
\end{aligned}
\end{equation*}
where $\ptvec(\PW)$ is reconstructed from the vectorial sum of \ptvecmiss and the lepton $\ptvec$ and $\ET(\PW)$ is calculated from the scalar sum of the lepton transverse energy and $\ptmiss$. Variables such as the invariant mass of the leptonically decaying $\W\Z$ system using constraints on the neutrino momentum from the $\W$ boson mass~\cite{Barr:2011xt} may be explored in future analyses.

A combination of methods using control samples in the data and detailed simulation studies is used to estimate background contributions. The following background categories are considered: $\W\Z$, $\Z\Z \to 4\ell$, $\V\V\V$, $\Z\gamma$, and processes with nonprompt leptons.

The QCD and EW $\W\Z$ background constitutes about 80\% of the total expected SM background yield.
The normalization of the QCD $\W\Z$ background is obtained from a background-dominated sideband, outside of the search region and defined by the dijet variables, where the expected signal yield is negligible: $100\GeV< m_{jj} < 500$\GeV and $\abs{\Delta \eta_{jj}} < 2.5$.
In this phase-space region, expected background contributions from EW $\W\Z$, $\Z\Z \to 4\ell$, $\V\V\V$, $\Z\gamma$ production, and nonprompt leptons are estimated to contribute about 40\% to the yield and are subtracted from the overall 266 events observed in data. The simulated sample of QCD $\W\Z$ processes is then normalized to match the observed number of events in this control region.
The estimated normalization of events is consistent with the SM prediction obtained using the \POWHEG NLO cross-section calculation.
The EW $\W\Z$ background contributes about 30\% to the overall $\W\Z$ background processes in the signal region.

The $\Z\Z \to 4\ell$, $\V\V\V$, and $\Z\gamma$ contributions are estimated from simulated samples, with corrections to the lepton reconstruction, trigger and selection efficiencies, and momentum scale and resolution, estimated from data control samples. The overall expected contribution from these processes to the total background yield is about 10\%, and the uncertainties in the estimates are dominated by the statistical component introduced by the number of simulated events passing the event selection requirements.
The $\Z\Z \to 4\ell$ background is largely reduced by the $\ptmiss$ requirement and the veto on events containing an additional lepton.

The main contributions to nonprompt leptons are from $\cPZ$+jets and top-quark ($\ttbar$ and $\PQt\PW$) events, where at least one of the jets or a jet constituent is misidentified as an isolated lepton. The dominant background at the final-selection level is $\cPZ$+jets. According to the simulation, fewer than 10\% of the background events with at least one nonprompt lepton come from top-quark processes. Data control samples are used to estimate this background. Lepton candidates selected with loose identification requirements are defined in a sample of events dominated by dijet production. The efficiency for candidates to pass the full lepton selection criteria is measured and is parametrized as a function of $\pt$ and $\eta$. The calculated efficiencies are used as weights to extrapolate the yield of the sample of loose leptons to the sample of fully selected leptons. The background estimation method is validated on a nonprompt lepton $\PW$+jets and $\ttbar$ enriched sample, selected by inverting the $\Z$ boson mass or $\PQb$-tagging criteria, where good agreement between the data and prediction is observed.

Uncertainties in the data-to-simulation scale factors applied to leptons in simulated samples result in an overall 4\% normalization uncertainty for backgrounds estimated from the simulation.
The experimental uncertainties in the lepton momentum scale and resolution, $\ptmiss$ modeling, and jet energy scale are applied in simulated events by smearing and scaling the relevant observables and propagating the effects to the kinematic variables used in the analysis, in particular $\mt$. Uncertainties in the lepton momentum scale and resolution are smaller than 1\% per lepton depending on the $\pt$ and $\eta$ of the lepton, and the effect on the yields at the analysis selection level is less than 1\%. The uncertainties in the jet energy scale and resolution result in a 5\% uncertainty in the signal yields. The uncertainty in the resolution of the $\ptmiss$ measurement is 10\%. Randomly smearing the measured $\ptmiss$ by one standard deviation of the resolution gives rise to a 5\% variation in the estimation of signal yields after the full selection.
Uncertainties of 2.3\% and 2.5\% are assigned to the integrated luminosity measurements in the years 2015 and 2016, respectively~\cite{LUMI1,LUMI3}.
The effect of higher-order corrections to the signal cross section in the GM model is taken from Ref.~\cite{Zaro:2002500}. The theoretical uncertainty is dominated by missing higher-order EW corrections estimated to be 7\%. Uncertainties in the signal acceptance due to PDF choice and renormalization and factorization scales are 2--3\% and less than 1\%, respectively, estimated using the LO signal samples.  Added in quadrature, the contributions result in an 8\% uncertainty in the normalization of the signal samples.

The uncertainty in the estimation of the expected number of QCD $\W\Z$ events is 12\%, which is estimated from the measured yields in the two-jet control region.
An uncorrelated uncertainty of 30\% is assigned on the normalization of $\W\Z$ events produced via EW processes, estimated from the largest bin-by-bin differences after varying the renormalization and factorization scales.
The total uncertainty in the prediction of the nonprompt background varies bin by bin in the $\mt$ distribution between 30\% and 80\%, dominated by the low number of nonprompt leptons passing the sideband selection.
A summary of the relative systematic uncertainties in the estimated signal and background
yields is shown in Table~\ref{tab:HwzSystematics}.
\begin{table*}[htbp]
\topcaption{Relative systematic uncertainties in the estimated signal and background
    yields, in units of percent.\label{tab:HwzSystematics}}
  \centering
\cmsTableResize{\begin{scotch}{lcccccc}

 Source & Signal & $\W\Z$ & $\V\V\V$  & $\Z\gamma$ & $\Z\Z$ & Nonprompt    \\

        \hline
Integrated luminosity 2015 (2016)      & 2.3 (2.5) &  \NA  &  2.3 (2.5)  &   2.3 (2.5)  & 2.3 (2.5) & \NA	 \\
Lepton efficiency           & 4.0  &  \NA  &  4.0  &   4.0  & 4.0 & \NA	 \\
Lepton momentum scale       & 1.0 & 1.0 &  1.0  &   1.0  & 1.0 &  \NA 	 \\
Jet momentum scale          & 5.0  & 10.0 &  6.0  &   30.0  & 13.0 &  \NA 	 \\
$\ptmiss$ resolution           & 5.0 & 1.7 &  1.0  &   \NA  & 7.0 &  \NA 	 \\
$\PQb$ tagging                   & 2.0 &  \NA  &  2.0  &   2.0  & 2.0 &  \NA 	 \\
QCD (EW) $\PW\Z$ bkg. normalization   & \NA & 12 (30) &  \NA  &   \NA  & \NA &  \NA 	 \\
Nonprompt bkg. normalization    & \NA & \NA &  \NA  &   \NA  & \NA & 30--80	 \\
GM uncertainties           & 8 & \NA &  \NA  & \NA  & \NA &  \NA 	 \\
      \end{scotch}}
\end{table*}

\begin{figure}[htbp]
\centering
\includegraphics[width=\cmsFigWidth]{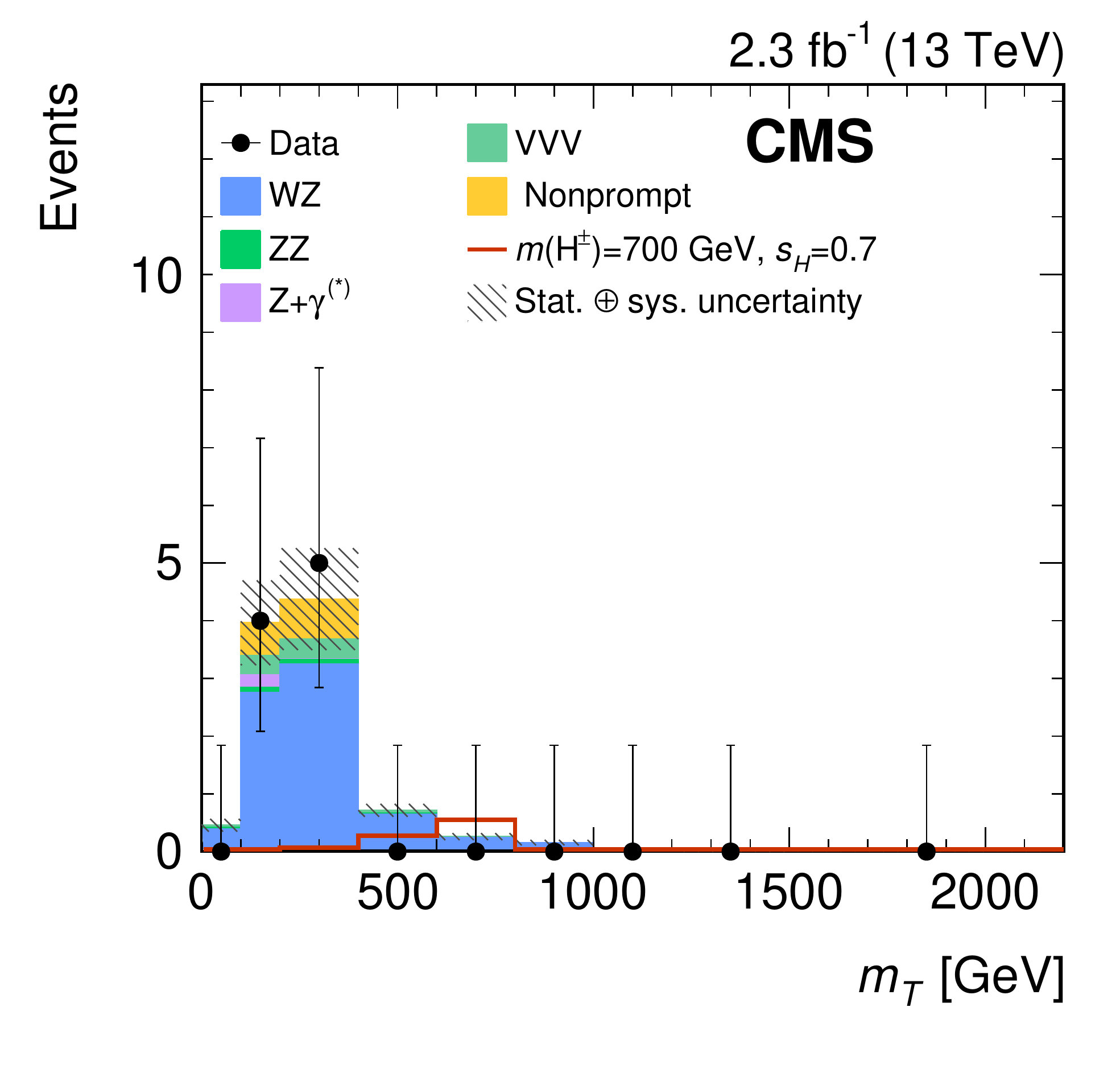}
\includegraphics[width=\cmsFigWidth]{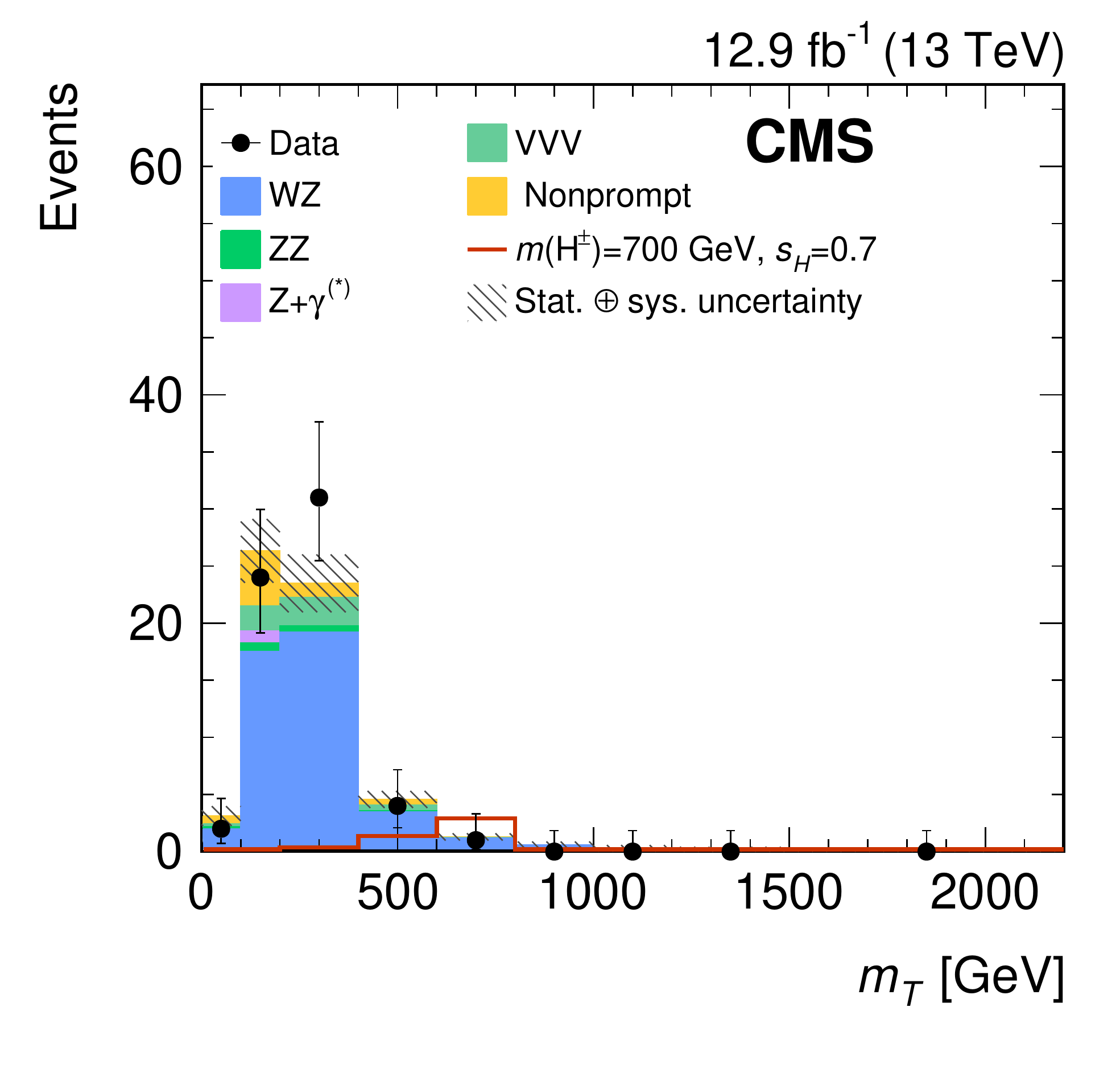}
\caption{Transverse mass distributions after full selection, for data collected in 2015 (\cmsLeft) and 2016 (\cmsRight). The background yield predictions correspond to the background-only hypothesis fit result. The signal distribution is shown for $m(\PHpm) = 700$\GeV and the cross-section prediction in the GM model at $s_{\PH}=0.7$.
}
\label{fig:distributions1_t}
\end{figure}

After applying the full selection, nine and 62 events are selected in the data collected in 2015 and 2016, respectively. The data yield together with the SM expectation for the different processes is given in Table~\ref{tab:sel_yields15}. The distribution of the $\mt$ with bin boundaries given by $\mt = [0,100,200,400,600,800,1000, 1200, 1500, \infty)$\GeV  (the last bin is an overflow bin) is shown in Fig.~\ref{fig:distributions1_t}. No event with $\mt(\PW\Z) >  800$\GeV is observed in the data, and overall agreement between the data and SM background prediction is observed.

\begin{figure}[htbp]
\centering
\includegraphics[width=\cmsFigWidth]{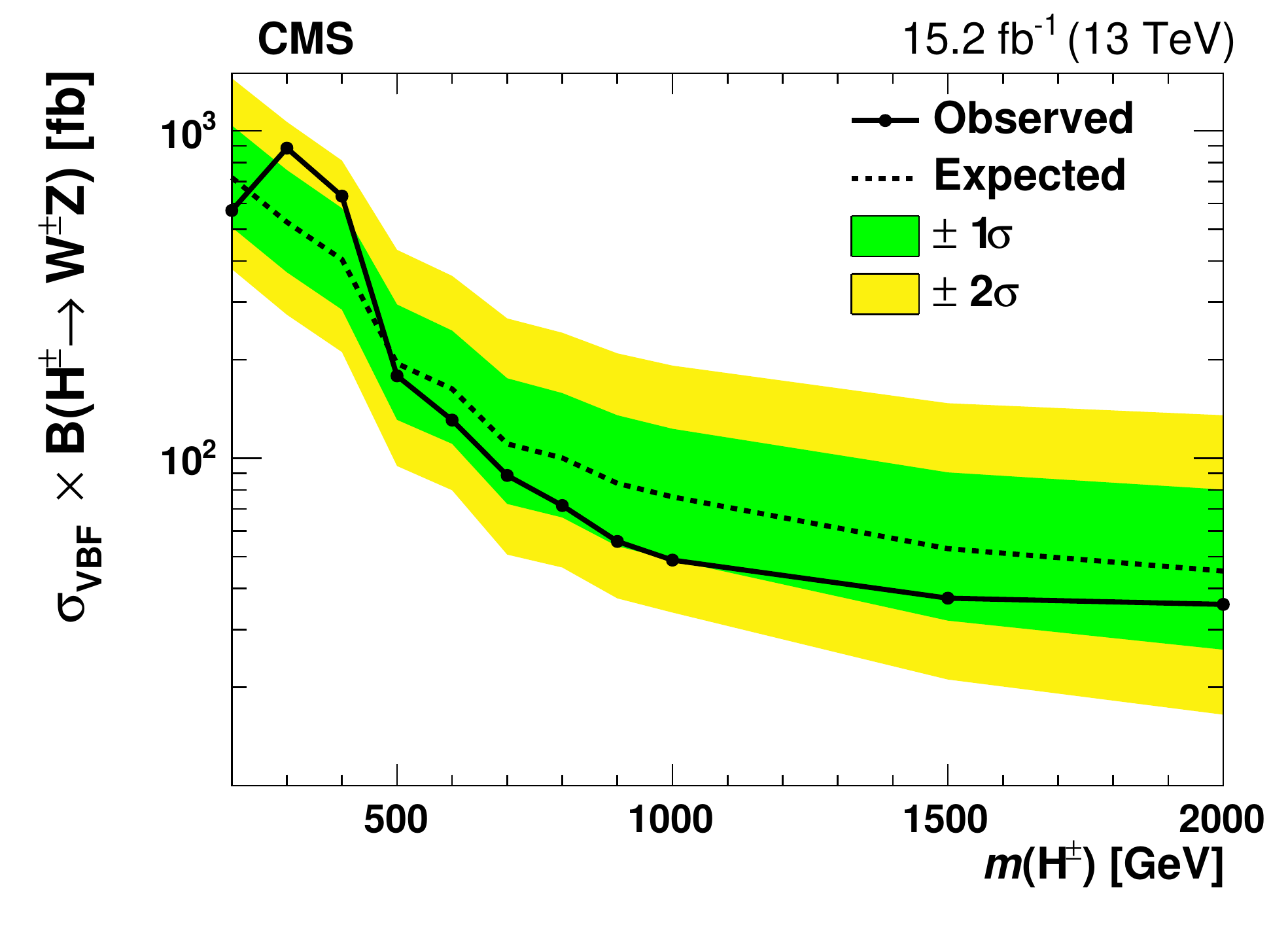}
\includegraphics[width=\cmsFigWidth]{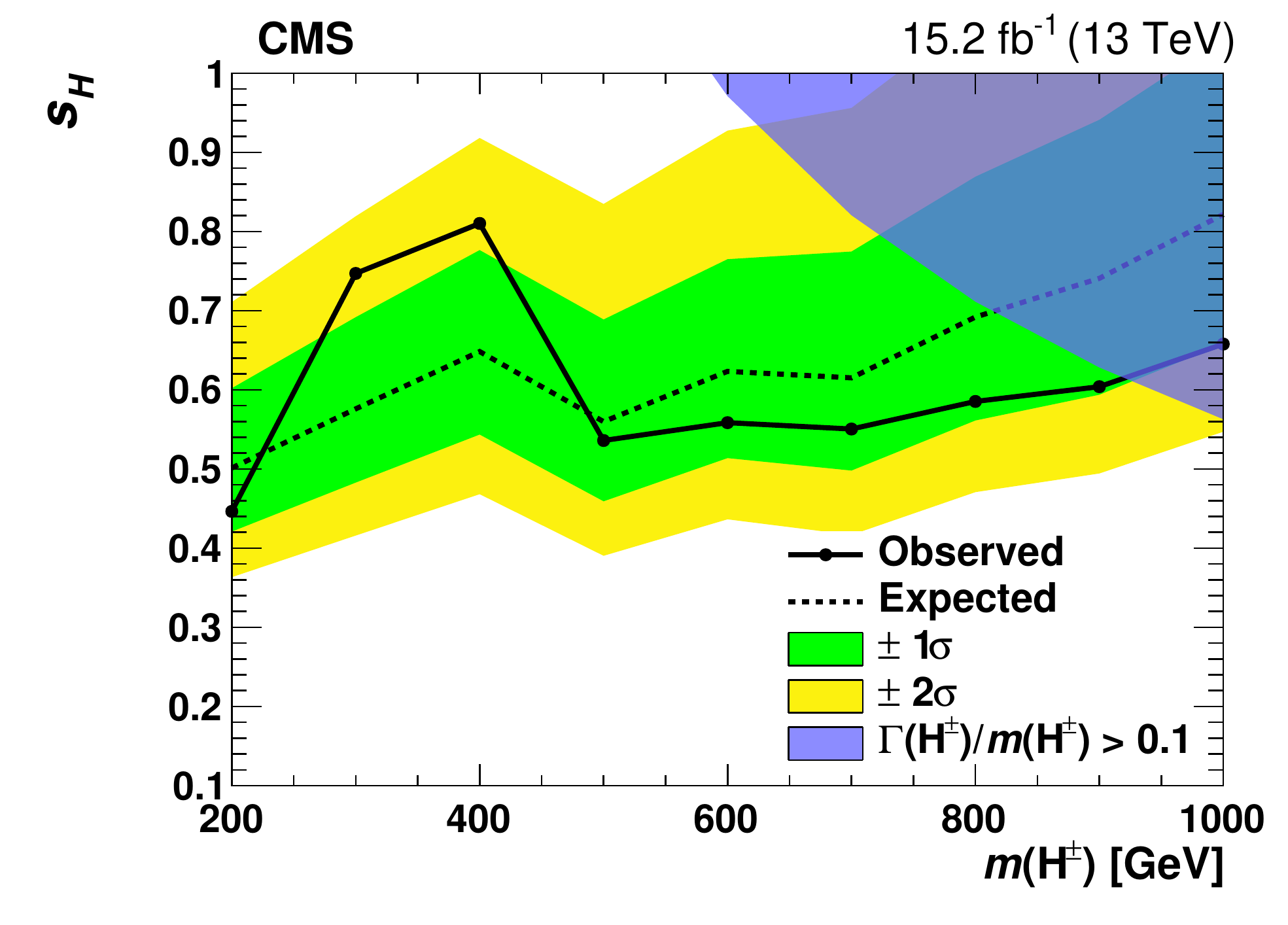}
\caption{Expected and observed exclusion limits at 95\% confidence
level as a function of $m(\PHpm)$ for $\sigma_\mathrm{VBF}(\PHpm) \, \mathcal{B}(\PHpm\to \PW\Z)$ (\cmsLeft) and on the ratio of vacuum expectation values in the GM model (\cmsRight) for 15.2\fbinv of proton-proton collisions at 13\TeV collected in 2015 and 2016. The blue shaded area covers the theoretically not allowed parameter space~\cite{Zaro:2002500}.
}
\label{fig:limits}
\end{figure}

A combined fit of the predicted signal and background yields in bins of $\mt$ to the data is performed to derive expected and observed exclusion limits on $\sigma_\mathrm{VBF}(\PHpm) \, \mathcal{B}(\PHpm\to \PW\Z)$ at 95\% confidence level using the CL$\mathrm{_s}$ method~\cite{junkcls,Read1,Cowan:2010js}. The exclusion limits as a function of $m(\PHpm)$, assuming a small intrinsic width for $\PHpm$, are shown in Fig.~\ref{fig:limits} (left). Values for $\sigma_\mathrm{VBF}(\PHpm) \,\mathcal{B}(\PHpm\to \PW\Z)$ ranging from 573\unit{fb} at $m(\PHpm) = 200$\GeV  to 36\unit{fb} at $m(\PHpm) = 2000$\GeV  are excluded by the data.

The model-independent exclusion limits are compared to the predicted cross sections at NNLO in the GM model~\cite{Zaro:2002500} in the $s_{\PH}$-$m(\PHpm)$ plane. For the probed parameter space and $\mt$ distribution used for signal extraction, the varying width as a function of $s_{\PH}$ is assumed to have negligible impact on the result. The value of the branching fraction $\mathcal{B}(\PHpm\to \PW\Z)$ is assumed to be one. In Fig.~\ref{fig:limits} (\cmsRight), the excluded $s_{\PH}$ values as a function of $m(\PHpm)$ are shown. The blue shaded region shows the parameter space for which the $\PHpm$ total width exceeds 10\% of $m(\PHpm)$, where the model is not applicable due to perturbativity and vacuum stability requirements~\cite{Zaro:2002500}. The observed limit excludes $s_{\PH}$ values greater than 0.45, 0.81, and 0.66 at $m(\PHpm) = 200$, 400, and 1000\GeV, respectively.

\begin{table}[htbp]
\topcaption{Yields of selected events in 2015 and 2016 data, together with the expected
yields from various background processes.  The statistical and systematic uncertainties are shown. The signal yields are shown for values of $s_{\PH}=0.7$.
\label{tab:sel_yields15}}
  \centering
  \newcolumntype{x}{D{,}{\,\pm\,}{3.3}}
  \begin{scotch}{lx{c}@{\hspace*{5pt}}x}
Data set   	      &  \multicolumn{1}{c}{2015} && \multicolumn{1}{c}{2016}	       	    \\
\cline{2-2}\cline{4-4}
Data                  &   \multicolumn{1}{c}{9}   && \multicolumn{1}{c}{62}               \\
\hline
$\W\Z$    	      &   7.5 ,  1.2  &&    44.4 ,  5.7    \\
$\Z\Z$    	      &   0.2 ,  0.1  &&    1.6 ,  0.2     \\
$\V\V\V$       	      &   0.8 ,  0.2  &&    5.5 ,  0.9   \\
$\Z\gamma$	      &   0.2 ,  0.1  &&    1.0 ,  0.6      \\
Nonprompt 	      &   1.3 ,  1.0  &&    7.4 ,  5.4     \\
\hline
Total bkg.           &    10.0 ,  1.6   &&    59.9 ,  8.0   \\[1ex]
Signal ($m(\PHpm) = 700$\GeV)    &  0.9 , 0.1 && 4.7 , 0.5\\
\end{scotch}
\end{table}


In summary, we present a search for charged Higgs bosons produced via vector boson fusion and decaying into $\PW$ and $\Z$ bosons in proton-proton collisions at $\sqrt{s} = 13$\TeV based on a sample corresponding to an integrated luminosity of 15.2\fbinv.
Events are required to have three leptons (electrons or muons), two jets
with large pseudorapidity separation and high dijet mass, and missing transverse momentum. The number of events observed in the signal region agrees with the standard model prediction. The first limits on $\sigma_\mathrm{VBF}(\PHpm)\, \mathcal{B}(\PHpm\to \PW\Z)$ at $\sqrt{s} = 13$\TeV are obtained. The results are interpreted in the Georgi--Machacek model for which the most stringent limits to date are derived.

We congratulate our colleagues in the CERN accelerator departments for the excellent performance of the LHC and thank the technical and administrative staffs at CERN and at other CMS institutes for their contributions to the success of the CMS effort. In addition, we gratefully acknowledge the computing centers and personnel of the Worldwide LHC Computing Grid for delivering so effectively the computing infrastructure essential to our analyses. Finally, we acknowledge the enduring support for the construction and operation of the LHC and the CMS detector provided by the following funding agencies: BMWFW and FWF (Austria); FNRS and FWO (Belgium); CNPq, CAPES, FAPERJ, and FAPESP (Brazil); MES (Bulgaria); CERN; CAS, MoST, and NSFC (China); COLCIENCIAS (Colombia); MSES and CSF (Croatia); RPF (Cyprus); SENESCYT (Ecuador); MoER, ERC IUT and ERDF (Estonia); Academy of Finland, MEC, and HIP (Finland); CEA and CNRS/IN2P3 (France); BMBF, DFG, and HGF (Germany); GSRT (Greece); OTKA and NIH (Hungary); DAE and DST (India); IPM (Iran); SFI (Ireland); INFN (Italy); MSIP and NRF (Republic of Korea); LAS (Lithuania); MOE and UM (Malaysia); BUAP, CINVESTAV, CONACYT, LNS, SEP, and UASLP-FAI (Mexico); MBIE (New Zealand); PAEC (Pakistan); MSHE and NSC (Poland); FCT (Portugal); JINR (Dubna); MON, RosAtom, RAS and RFBR (Russia); MESTD (Serbia); SEIDI and CPAN (Spain); Swiss Funding Agencies (Switzerland); MST (Taipei); ThEPCenter, IPST, STAR and NSTDA (Thailand); TUBITAK and TAEK (Turkey); NASU and SFFR (Ukraine); STFC (United Kingdom); DOE and NSF (USA).

\bibliography{auto_generated}

\cleardoublepage \appendix\section{The CMS Collaboration \label{app:collab}}\begin{sloppypar}\hyphenpenalty=5000\widowpenalty=500\clubpenalty=5000\textbf{Yerevan Physics Institute,  Yerevan,  Armenia}\\*[0pt]
A.M.~Sirunyan, A.~Tumasyan
\vskip\cmsinstskip
\textbf{Institut f\"{u}r Hochenergiephysik,  Wien,  Austria}\\*[0pt]
W.~Adam, E.~Asilar, T.~Bergauer, J.~Brandstetter, E.~Brondolin, M.~Dragicevic, J.~Er\"{o}, M.~Flechl, M.~Friedl, R.~Fr\"{u}hwirth\cmsAuthorMark{1}, V.M.~Ghete, C.~Hartl, N.~H\"{o}rmann, J.~Hrubec, M.~Jeitler\cmsAuthorMark{1}, A.~K\"{o}nig, I.~Kr\"{a}tschmer, D.~Liko, T.~Matsushita, I.~Mikulec, D.~Rabady, N.~Rad, B.~Rahbaran, H.~Rohringer, J.~Schieck\cmsAuthorMark{1}, J.~Strauss, W.~Waltenberger, C.-E.~Wulz\cmsAuthorMark{1}
% ...
\vskip\cmsinstskip
% ...
1:~~Also at Vienna University of Technology, Vienna, Austria\\
% ...

\end{sloppypar}
\end{document}